\documentstyle[12pt]{article}

\begin{document}

{\it University of Shizuoka}

\hspace*{9.5cm} {\bf US-95-03}\\[-.3in]

\hspace*{9.5cm} {\bf AMU-95-04}\\[-.3in]

\hspace*{9.5cm} {\bf Revised Version}\\[-.3in]

\hspace*{9.5cm} {\bf October 1995}\\[.1in]


\begin{center}

{\large\bf  Top-Quark-Mass Enhancement }\\[.1in]

{\large\bf  in a Seesaw-Type Quark Mass Matrix }\footnote{
hep-ph/9505201: To be published in Z.~Phys.~C.}\\[.2in]

{\bf Yoshio Koide}\footnote{
E-mail: koide@u-shizuoka-ken.ac.jp} \\

Department of Physics, University of Shizuoka \\
395 Yada, Shizuoka 422, Japan \\[.05in]

and \\[.05in]

{\bf Hideo Fusaoka}\footnote{
E-mail: fusaoka@amugw.aichi-med-u.ac.jp} \\

Department of Physics, Aichi Medical University \\
Nagakute, Aichi 480-11, Japan \\[.2in]

{\large\bf Abstract}\\[.1in]

\end{center}

\begin{quotation}
We investigate the implications of a seesaw type mass matrix, i.e.,
$M_f\simeq m_L M_F^{-1} m_R$,  for quarks and leptons $f$
under the assumption that
the matrices $m_L$ and $m_R$ are common to all flavors (up-/down-
and quark-/lepton- sectors)
and the matrices $M_F$ characterizing the heavy fermion sectors
have the form [(unit matrix)
+ $b_f$ (a democratic matrix)] where $b_f$ is a flavor parameter.
We find that by adjusting the complex parameter $b_f$,
the model can provide that
$m_t\gg m_b$ while at the same time keeping $m_u\sim m_d$
without assuming any parameter with
hierarchically different values between $M_U$ and $M_D$.
The model with three adjustable parameters
under the ``maximal" top quark mass enhancement can give
reasonable values of five quark mass ratios and
four KM matrix parameters.
\end{quotation}

\newpage

\noindent
{\large\bf 1. Introduction}

One of the most mysterious facts in the quark mass spectrum is
why top quark mass $m_t$ is so much larger than the bottom quark mass $m_b$,
while $u$ quark mass $m_u$ is of the order of $d$ quark mass $m_d$.
In the usual discussion of fermion masses, this drastic generation dependence
of the mass splitting between members of each isomultiplet of quarks
is  attributed to an arbitrary hierarchy among the input
parameters which is not completely satisfactory. It is therefore
important to seek alternative ways to understand this feature.
In this paper we argue that within the see-saw[1] type mass formula for
quark masses discussed in the context of gauge models [2],
a very simple explanation of this feature is obtained
by imposing a specific universality ansatz for various flavor matrices.
 We then find that
a slight generalization of this ansatz provides an extremely good fit
to all the quark mass ratios and mixings.

Our starting point is the following specific see-saw type ansatz
proposed by one of the authors [3] for
  quark and lepton mass matrices:
$$
M_f = M_e^{1/2} O_f M_e^{1/2} \ , \eqno(1.1)
$$
where $M_e^{1/2}={\rm diag}(\sqrt{m_e}, \sqrt{m_\mu}, \sqrt{m_\tau})$.
Here, for the up-quark mass matrix $M_u$, the matrix $O_f$ ($f=u$) is given
by
$$
O_f={\bf 1} + 3 a_f X \ , \eqno(1.2)
$$
where {\bf 1} is a unit matrix and $X$ is a democratic-type matrix
[4]
$$
X=\frac{1}{3} \left(
\begin{array}{ccc}
1 & 1 & 1 \\
1 & 1 & 1 \\
1 & 1 & 1
\end{array} \right) \ , \eqno(1.3)
$$
which satisfies the relation $X^2=X$.
The up-quark mass matrix  can then successfully give
a quark mass ratio [3,5]
$$
\frac{m_u}{m_c} \simeq \frac{3m_e}{4m_\mu} \ , \eqno(1.4)
$$
for a large value of the parameter $a_u$.
The value of $a_u$ is adjusted from the mass ratio $m_c/m_t$ .

Stimulated by the phenomenological success of the mass matrix form
(1.1) -- (1.3),
the authors [6] have applied the mass matrix form to down-quark
mass matrix, by considering that the parameter $a_d$ is complex.
They have found that the value of a complex parameter $a_d$
which fits the mass ratios $m_d/m_s$ and $m_s/m_b$ gives
reasonable values of not only Kobayashi-Maskawa (KM) [7]
matrix elements $V_{ij}$ ($i,j$ denote family indices)
but also up-to-down-quark mass ratios
$m_u/m_d$, $m_c/m_s$ and $m_t/m_b$.

Suggested from the form (1.1), it may be expected that
such  phenomenological success will also
be obtained in the context of a seesaw-type mass matrix
$$
M_f\simeq m_L M_F^{-1} m_R \ , \eqno(1.5)
$$
with $m_L\propto m_R \propto M_e^{1/2}$
and $M_F \propto O_f^{-1}$.
Here, the expression (1.5) is derived from the
$6\times 6$ mass matrix for fermions $(f, F)$
$$
(\overline{f}_L\ \overline{F}_L) \left(
\begin{array}{ccc}
0 & m_L \\
m_R & M_F
\end{array} \right) \left(
\begin{array}{c}
f_R \\
F_R
\end{array} \right) + h.c. \ , \eqno(1.6)
$$
for the case of $O(M_F)\gg O(m_R), O(m_L)$,
where$f=(f_1,f_2,f_3)$ are three family
quarks and leptons, and $F=(F_1,F_2,F_3)$ are
vector-like heavy fermions corresponding to $f$.

The re-interpretation of the model (1.1) based on the seesaw
model (1.5) seems to be plausible because of the
following reasons.
The inverse of the matrix $O_f$ with a simple form
[(unit matrix)$+$(democratic-type matrix)] has also
a simple form
[(unit matrix)$+$(democratic-type matrix)], i.e.,
$$
O_F\equiv O_f^{-1}={\bf 1} + 3 b_f X \ , \eqno(1.7)
$$
where the complex coefficients $a_f$ and $b_f$ are related by
$$
a_f  =- b_f /(1+3b_f ) \ . \eqno(1.8)
$$
In the mass matrix model (1.1), we need hierarchically different
values  [6] of
the parameters $a_f$, i.e., $a_u=28.65$ and $|a_d|=0.4682$,
in order to provide reasonable quark masses and KM mixings,
while, as seen from (1.8), the values $|a_u|\gg 1$ and
$a_d\simeq -1/2$ correspond to $b_u\simeq -1/3$ and $b_d\simeq -1$ in
the inverse matrix (1.7), respectively.
In the present paper, we are interested in such a model
that $M_u$ and $M_d$ are ``almost"
symmetric, i.e., they have almost the same structure and
they take parameter values which are not so hierarchically different
between $M_u$ and $M_d$.
The parameter ratio $|a_u/a_d|\simeq 60$ in the model (1.1) can be reduced
to the ratio $|b_d/b_u| \simeq 3$ in (1.7).

However, when we consider a model (1.6) (not (1.5)) with
$M_F \propto O_F$, one problem arises:
Recently the CDF collaboration [8] has  reported
$m_t=174\pm 10^{+13}_{-12}$ GeV as top quark mass from $\overline{p}p$
collision data at $\sqrt{s}=1.8$ TeV.
On the other hand, the universal mass matrix $m_L$ which breaks the
SU(2)$_L$ gauge symmetry should be  of the order
$\Lambda_W =(\sqrt{2}G_F)^{-1/2}/\sqrt{2}=174$ GeV$\sim m_t$, or less.
Then, the approximate expression (1.4) for up-quarks is not valid
any longer,
because if (1.5) is valid,  $O(m_L)  \sim m_t$ means
$M_U^{-1}m_R \sim O(1)$,
so that it does not satisfy the condition $O(M_F)\gg O(m_R)$ for the
validity of the seesaw expression (1.5).
This is also understood from the fact
that the limit $|a_u|\rightarrow\infty$ means
the limit $b_u\rightarrow -1/3$ and the determinant of $M_U$
becomes zero in the limit, so that the expansion of $M_f$
in $M_F^{-1}$ can not be a good approximation.

In this paper, we do not use the approximate relation (1.5),
but  calculate directly the $6\times 6$ mass matrix (1.6).
In Sect.~2, we will give the outline of our mass matrix model.
In Sect.~3, we will give an expression of $M_f$ which is
valid in the limit of $b_f\rightarrow -1/3$, i.e.,
det$\, M_F=0$, instead of the well-known seesaw expression (1.5),
and discuss the up-quark mass ratios which are expressed
in terms of lepton mass ratios and our adjustable parameters
(see the next section).
In Sec.~4, we discuss the fermion mass spectra
by numerically evaluating the $6\times 6$ mass matrix.
In Sect.~5, KM matrix parameters are discussed numerically.
In the present model, under some basic assumptions
(see Sects.~2 and 5),
the parameter fitting for quark mass ratios and KM matrix
parameters (5+4=9 observables) is done by three adjustable
parameters $k/K$, $b_d$ and $\beta_d$ (see the next
section for the definitions).
We will find that the value of $m_t$ takes the largest enhancement at
$b_u=-1/3$, while the relations  $m_u\sim m_d$ and (1.4) are kept.
We can obtain reasonable values of quark mass ratios (not only
$m_u/m_c$, $m_c/m_t$, $m_d/m_s$ and $m_s/m_b$, but also $m_u/m_d$,
$m_c/m_s$ and $m_t/m_b$) and the KM matrix parameters, by taking
$b_u=-1/3$ and $b_d \simeq -1$.

\vglue.2in

\noindent
{\large\bf 2. Outline of the model}

In addition to the conventional quarks and leptons $f_i$,
where $f$ is the flavor index ($f=u, d, \nu$ and $e$ denote
up-quarks, down-quarks, neutrinos and charged leptons), and
$i$ is the family index ($i=1,2,3$), We consider vector-like fermions
$F_i$ correspondingly to $f_i$.
These fermions belong to $f_L=(2,1)$, $f_R=(1,2)$,
$F_L=(1,1)$ and $F_R=(1,1)$ of SU(2)$_L\times$SU(2)$_R$.
A ``would-be" seesaw mass matrix for the fermions $(f,F)$
is given by (1.6).
Gauge models
which  realize the mass matrix form (1.6) have been proposed
by many authors [2].
Although the interest of most authors is  how to embed
the model (1.6) into a unification model in the framework of
gauge theory, our interest is how to give realistic quark
mass spectra and family mixing from the phenomenological
point of view.

Suggested by the phenomenological success of the model (1.1),
we assume the following mass matrix [9]
$$
M=\left(
\begin{array}{ccc}
0 & m_L \\
m_R & M_F
\end{array} \right)
= m_0 \left(
\begin{array}{ccc}
0 & Z \\
k Z & K O_F
\end{array} \right) \ , \eqno(2.1)
$$
where the matrices $m_L$ and $m_R$ (i.e., $m_0$, $h$ and the matrix $Z$)
are common to all of $f=u,d,\nu,e$, and only $M_F$ depends
on flavors $f$ through the complex parameter $b_f$.
Hereafter, we denote the complex parameter $b_f$ in (1.7) as
$b_f e^{i\beta}$ ($b_f$ is real and $|\beta_f|\leq \pi/2$) in (2.2) below.
The vector-like fermions $F$ acquire large masses $M_F$
at an energy scale $\mu=m_0 K$.
We consider that the energy scale $m_0 K$ is not
as large as the ground unification scale, but an
intermediate energy scale.
At the present stage, the origin of the democratic form
$$
O_F={\bf 1} + 3 b_f e^{i\beta_f} X
= \left( \begin{array}{ccc}
1 & 0 & 0 \\
0 & 1 & 0 \\
0 & 0 & 1
\end{array} \right) + b_f e^{i\beta_f}  \left(
\begin{array}{ccc}
1 & 1 & 1 \\
1 & 1 & 1 \\
1 & 1 & 1
\end{array} \right) \ , \eqno(2.2)
$$
is an open question.
We may attribute the origin of the democratic term $X$
to a permutation symmetry S$_3$ [10], a BCS-like
mechanism [11], a composite model based on the analogy
of hadronic $\pi^0$-$\eta$-$\eta'$ mixing [12], and
so on.
In the present phenomenological analysis,
we do not discuss its origin moreover.

The present model is left-right symmetric except for
$k\neq 1$. At an energy scale $\mu=m_0 k$ ($\mu=m_0$)
at which SU(2)$_R$ (SU(2)$_L$) is broken, the mass term
$\overline{F}_L m_R f_R$ ($\overline{f}_L m_L F_R$)
appears, so that we consider $k \sim m(W_R)/m(W_L)$.
The relation $m_L=m_R/k=m_0 Z$ is merely a
phenomenological working hypothesis.
The matrix $Z$ takes a diagonal form
$$
Z={\rm diag} (z_1, z_2, z_3) \ , \eqno(2.3)
$$
with the normalization condition
$z_1^2+z_2^2+z_3^2=1$. (In other words, in the
family basis in which $Z$ is diagonal, we have assumed
that the matrix $O_F$ is given by (2.2)).
For the charged leptons, since $m_\tau \ll m_0
\sim m_W$, it is clear that the seesaw expression
$M_e=m_0 (k/K) Z O_F^{-1}Z$
is well satisfied, so that we can fix the parameter
$z_i$ as
$$
\frac{z_1}{\sqrt{m_e}}=\frac{z_2}{\sqrt{m_\mu}}
=\frac{z_3}{\sqrt{m_\tau}}=\frac{1}{\sqrt{m_e+m_\mu+m_\tau}}
\ . \eqno(2.4)
$$
Here, we have assumed $b_e=0$ according to the
phenomenological success [3] of the model (1.1).
In the present paper, we do not discuss why $z_i$ are
given by the relation (2.4),
because the purpose of the present paper is to study
quark mass ratios and KM matrix parameters
phenomenologically, so that
charged lepton masses are regarded as inputs
in the numerical estimates.
Since the evolution effects of fermion mass ratios (not
the absolute values) from $\mu=m_0 K$ to $\mu=m_0$
are, at most, several percent, for simplicity,
we use the values of
$z_i$ which are fixed by using the formula (2.4)
with the observed charged lepton masses [13].

For the case of $K\gg k\gg 1$, the quark mass
ratios and the KM matrix parameters (nine observables)
are described by five real parameters
$k/K$ (not $k$ and $K$ separately), $b_u$, $\beta_u$, $b_d$
and $\beta_d$.
As we will discuss in Sections 3 and 4,
the maximal top-quark-mass enhancement occurs
at $b_u=-1/3$ and $\beta_u=0$.
We will put an ansatz of ``maximal top-quark-mass
enhancement", so that we will fix the parameters $b_u$
and $\beta_u$ to $b_u=-1/3$ and $\beta_u=0$.
The numerical fitting for the nine observables is then tried
by adjusting only three parameters $k/K$, $b_d$ and
$\beta_d$.
However, as will be discussed in Sect.~5, a straightforward
application of the mass-matrix model (2.1) cannot lead to
reasonable predictions of the KM matrix parameters.
We will therefore introduce a sign factor by replacing
$m_L=m_0 Z$ in (2.1) by $m_L^f=m_0 P_f Z$,
where $P_u={\rm diag}(1,1,1)$, while $P_d={\rm diag}(1,1,-1)$.
The adjustable parameters are still three, i.e.,
$k/K$, $b_d$ and $\beta_d$.
The phase matrices $P_f$ do not affect the discussion of
the mass spectrum.
For a time being in Sects.~3 and 4, we will neglect the phase
matrices $P_f$.

\vglue.2in

\noindent
{\large\bf 3. \  Expression of $M_f$ in the case of $b_f\simeq -1/3$}

One of the purposes in the present paper is to obtain a reliable
expression of $M_f$ in the case of $b_f\simeq -1/3$, because
the case leads to det$\, M_F\simeq 0$, so that the seesaw expression
(1.5) which is obtained by expanding it in $M_F^{-1}$
is not valid any longer.

As shown in Appendix, in general, the transformation of the
$6\times 6$ mass matrix $M$ into
$$
U_L M U_R^\dagger
\equiv U_L \left(
\begin{array}{cc}
M_{11} & M_{12} \\
M_{21} & M_{22}
\end{array} \right) U_R^\dagger
=M'\equiv  \left(
\begin{array}{cc}
M_{11}^{'} & 0 \\
0 & M_{22}^{'}
\end{array} \right) \ ,
\eqno(3.1)
$$
is done by the following
two $6\times 6$ unitary matrices,
$$ U_L = \left(
\begin{array}{cc}
(1+\rho_L\rho_L^\dagger)^{-1/2} & (1+\rho_L\rho_L^\dagger)^{-1/2}\rho_L \\
-(1+\rho_L^\dagger\rho_L)^{-1/2}\rho_L^\dagger &
(1+\rho_L^\dagger\rho_L)^{-1/2}
\end{array} \right) \  \eqno(3.2)
$$
and $U_R$ with $L\leftrightarrow R$ in (3.2).
The so-called seesaw expression $M'_{11}\equiv M_f\simeq m_L M_F^{-1}m_R$
is obtained by expanding $M'_{11}$ in $M_F^{-1}$.
Since our mass matrix (2.1) is not Hermitian,
for evaluating the KM matrix
(family mixing of left-handed fermions),
it is useful to define
the $3\times 3$ Hermitian matrix $H_f$:
$$
H_f \equiv M'_{11}M_{11}^{\prime\dagger}
= (1+\rho_L\rho_L^\dagger)^{-1/2}\widetilde{H}_f
(1+\rho_L\rho_L^\dagger)^{+1/2} \ . \eqno(3.3)
$$
As seen in (A.22), (A.24) and (A.27),
the matrix $\widetilde{H}_f$ is given by
$$
\widetilde{H}_f
\equiv
\rho_L m_R \rho_R m_L^\dagger
= (m_L+\rho_L M_F) m_L^\dagger \ , \eqno(3.4)
$$
and it satisfies the following equation:
$$
\widetilde{H}_f^2 m_L^{\dagger -1}
- \widetilde{H}_f m_L^{\dagger -1}
\left(M_F^\dagger M_F +m_L^\dagger m_L
+M_F^{-1} m_R m_R^\dagger M_F \right)
+ m_L M_F^{-1} m_R m_R^\dagger M_F =0 \ . \eqno(3.5)
$$

Our interest is in the expression of $\widetilde{H}_f$
in the case of det$\,M_F\simeq 0$.
However, since it is hard to obtain the general formulation
in such the case,
we confine ourselves to investigating the special form (2.1)
with (2.2).

For the investigation of the case of $b_u\simeq -1/3$, it is convenient
to define the parameter
$$
3 \varepsilon \equiv \Delta b = b +\frac{1}{3} \ . \eqno(3.6)
$$
Then, the matrix $O_F$ is represented by
$$
O_F=Y+\varepsilon X \ , \eqno(3.7)
$$
where
$$
Y={\bf 1} - X \ , \eqno(3.8)
$$
and the matrices $X$ and $Y$ satisfy the relations
$X^2=X$,  $Y^2=Y$, and $XY=YX=0$ from the definitions (1.3) and (3.8),
so that the inverse of $O_F$, (3.7),
is given by
$$
O_F^{-1}=Y+X/\varepsilon \ . \eqno(3.9)
$$

For the case of ($k/K)^2 \ll \varepsilon^2 \ll 1$,
from the equation (3.5), we obtain
$$
\widetilde{H}_f \simeq m_0^2 \left(\frac{k}{K}\right)^2
Z\left(Y+\frac{1}{\varepsilon}X\right)Z^2
\left(Y+\frac{1}{\varepsilon}X\right)Z \ , \eqno(3.10)
$$
which corresponds to the well-known seesaw expression
$M_f\simeq m_0 (k/K) Z O_F^{-1} Z$.

For a general case,
we assume an approximate form
$$
\widetilde{H}_u \simeq m_0^2  Z \left(\frac{k}{K}Y+xX\right)Z^2
\left(\frac{k}{K}Y+xX\right)Z \ , \eqno(3.11)
$$
from an analogy to the form (3.10).
By substituting (3.11) into (3.5), we find
$$
x \simeq \left[\frac{\varepsilon}{2k/K} + \sqrt{
\frac{1}{3} + \left(\frac{\varepsilon}{2k/K}\right)^2}
\  \right]^{-1} \ . \eqno(3.12)
$$
For $\varepsilon^2 \gg (k/K)^2$, (3.12) reproduces (3.10).
For $\varepsilon^2 \ll (k/K)^2$, we obtain
$$
\widetilde{H}_u \simeq 3 m_0^2 Z \left(X+ \frac{1}{\sqrt{3}}\frac{k}{K}Y
\right)Z^2
\left(X+ \frac{1}{\sqrt{3}}\frac{k}{K}Y\right)Z \ . \eqno(3.13)
$$
This expression (3.13) is the expression which should be used
in the case of det$\, M_F\simeq 0$ as a substitute
for the well-known seesaw expression
(3.10).

The mass eigenvalues are calculated from
${\rm Tr}{H}_u ={\rm Tr}\widetilde{H}_u$,
$(({\rm Tr}{H}_u)^2- {\rm Tr}{H}_u^2)/2= (({\rm Tr}\widetilde{H}_u)^2
- {\rm Tr}\widetilde{H}_u^2)/2$ and
$ {\rm det} {H}_u= {\rm det} \widetilde{H}_u$.
We obtain up-quark masses
$$
m_u \simeq \frac{3}{2}z_1^2 \frac{k}{K}m_0 \ , \ \ \
m_c \simeq 2z_2^2z_3^2\frac{k}{K} m_0 \ , \ \ \
m_t \simeq \frac{1}{\sqrt{3}}\frac{1}{\sqrt{1+27(\Delta b)^2 (K/k)^2}}
m_0  \ ,
\eqno(3.14)
$$
from (3.11) and (3.12), where $\varepsilon=\Delta b/3$ (3.6).
We find that the relation (1.4) is also valid in the case
$(\Delta b)^2 \ll (k/K)^2 \ll 1$, even in the limit of $b_u=-1/3$.

\vglue.2in

\noindent
{\large\bf 4. \ Numerical study of quark mass ratios}

Numerical evaluation of the eigenvalues of the $6\times 6$
mass matrix (2.1) can  easily be done with the help of a computer.
Numerical study is helpful for checking analytical calculations
based on the formalism of the previous section.
In Fig.~1, in order to give an overview of the mass spectrum
in our mass matrix model, we illustrate
the light fermion mass spectrum $m_i^f$ ($i=1,2,3$) versus the
parameter $b_f e^{i\beta_f}$.
Here, we have taken $k=10$ and $K/k=50$ as a trial.
(The choices of $k$ and $K/k$ are discussed later.)
In order to fix the values of the parameters $z_i$ at $b_e=0$,
we have used the observed charged leptons masses [13] as inputs.

The spectrum for the case of $\beta_f=0$ (solid lines) shows
the following characteristics:

\noindent
(1) The third fermion mass is sharply enhanced at $b_f=-1/3$.

\noindent
(2) Level crossing (mass degeneration) occurs at
$b_f=-1/2$ and $b_f=-1$.

\noindent
These characteristics become mild when $\beta_f$ takes
a sizable value (dashed lines).

For comparison, we  list
 the observed running quark mass values
(in unit of GeV) [14]
at $\mu=\Lambda_W\equiv (\sqrt{2} G_F)^{-1/2}=174$ GeV:
$$
\begin{array}{lll}
m_u=0.00230 \pm 0.00045 \ , & m_c=0.612^{+0.010}_{-0.023} \ ,
& m_t=166^{+21}_{-26}\ , \\
m_d=0.00406 \pm 0.00045 \ , & m_s=0.082 \pm 0.014 \ ,
& m_b=2.874^{+0.012}_{-0.023} \ . \\
\end{array} \eqno(4.1)
$$

In the previous section, we have showed that
the up-quark mass ratio
$m_u/m_c$ is given by (1.4) in the limit $\varepsilon \ll (k/K)^2\ll 1$,
see (3.14).
The relation can be checked by a numerical study.
We find that
the ratio $m_c/m_u$ at a fixed $K/k$ is
insensitive to the choice of $k$,
for $k \geq 10$.
Also, the ratio is insensitive to
the parameters $K/k$  and $\Delta b_u$
for large $K/k$;
for example, $m_c/m_u=260.8, 260.8, 259.2$, and $259.2$,
for $(K, k, \Delta b_u) = (10^3, 10, 0), (10^5, 10, 0)$,
$(10^3, 10, 0.003)$ and $(10^5, 10, 0.003)$, respectively,
while ($m_c/m_u)_{exp} = 266_{-49}^{+70}$.
Thus, we conclude that the relation (1.4) is valid
almost independently of the  values of $k$ and $K/k$
for the case of $K\gg k \gg 1$.

Next, we study the up-quark mass ratio $m_t/m_c$.
We find that the ratio is also insensitive to the value of $k$
for $k\geq 10$.
Therefore, we illustrate the behavior of $m_t/m_c$ versus $K/k$
for the case of $k=10$  in Fig.~2.
It is noticeable that, for $\Delta b_u\simeq +0.00388$ and
$\Delta b_u\simeq -0.00362$, the ratio
$m_t/m_c$  comes near the experimental value
$\left(m_t/m_c\right)_{exp}\simeq 271$
as  $K/k\rightarrow \infty$.
For the case $|\Delta b_u|\geq 0.005$, we cannot fit the ratio
$m_c/m_t$ suitably, so that
the case is ruled out.
For $\Delta b_u=0$, the ratio $m_t/m_c$ increases linearly in $K/k$.
In order to fit the prediction to the experimental value of
$\left(m_t/m_c\right)_{exp}= 271\pm 46$,
we need $K/k= 50\pm 8$ for the case $\Delta b_u=0$ whereas we find
$K/k= (2.0^{+\infty}_{-1.3})\times 10^{2}$
for the cases $\Delta b_u=+0.00388$ and $\Delta b_u=-0.00362$.

Although a scenario with $\Delta b_u \simeq \pm 0.004$ and
$K/k >2\times 10^2$ seems to be attractive because the ratio $m_t/m_c$
can be fitted insensitive to $K/k$, we do not adopt this scenario
because of the following consideration of the absolute value of $m_t$.
In Fig.~3, we show the behavior of $m_t/m_0$ versus $K/k$.
Since the ratio is again insensitive to the value of $k$ for $k\geq 10$,
we illustrate the case of $k=10$.
In the limit of $b_u=-1/3$, the value $m_t/m_0$ is almost constant,
i.e.,  $m_t/m_0\simeq 1/\sqrt{3}$ (a) as we have shown in (3.14).
On the other hand, as seen in Fig.~3, the case (c) $\Delta b_u\simeq
\pm 0.004$ gives $m_t/m_0 < 0.161$ for $K/k >2\times 10^2$.
If we consider that the mass matrix $m_L$ originates from the couplings
to an SU(2)$_L$ doublet Higgs
boson  $\phi_{L}$  with the vacuum expectation value (VEV)
$\langle\phi_{L}^0\rangle_0=v_0=\Lambda_W=174$ GeV,
the Yukawa coupling constants $y_{Li}$  with
fermions $\overline{f}_{Li} F_{Ri}$ are given by
$ y_{Li}=z_i m_0/ v_0$.
Therefore, a small value of $m_t/m_0$ means a large value of
$y_{L3} = z_3 (m_0/m_t)(m_t/v_0)$.
The value $m_t/m_0=0.161$ corresponds to $y_{L3} = 6.03 (m_t/v_0)$.
Such a large value may be unfavorable from the point of view
of the perturbative electroweak theory.
Hereafter, we adopt the ansatz of the ``maximal top-quark-mass
enhancement", i.e., $b_u=-1/3$ (solid lines in Figs.~2 and 3),
and we fix the parameter $K/k$ to $K/k=50$
from the observed ratio of $m_t/m_c$.

On the other hand, the down-quark masses are given by adjusting
two parameters $b_d$ and $\beta_d$.
As seen in Fig.~1, the case of $b_d \simeq {-1}$ is favorable because
it can give reasonable predictions  not only for $m_b/m_s$
and $m_s/m_d$, but also for $m_d/m_u$.
The ratios $m_s/m_d$ and $m_b/m_s$ versus $b_d$ and $\beta_d$ are
illustrated in Fig.~4 for the case of $K/k=50$ and $k=10$.
As far as we see in Fig.~4, the cases $b_d=-1.1 \sim -1$ with
$\beta_d= -20^\circ \sim -16^\circ$
 are favorable.
Considering the present experimental uncertainty of quark mass values,
hereafter, we simply adopt the integral solution $b_d=-1$
for further numerical estimates.

For the case of $b_d\simeq -1$ and $1\gg\beta_d^2\neq 0$,
down-quark masses are given by
$$
m_d\simeq z_1^2 \frac{1}{\beta_d}\frac{k}{K}m_0 \ , \ \ \
m_s\simeq z_2^2 z_3^2 {\beta_d} \frac{k}{K}m_0 \ , \ \ \
m_b\simeq \frac{1}{2}\frac{k}{K}m_0 \ .
\eqno(4.2)
$$
In the present model, the up-to-down quark mass ratio
$m_u/m_d$ is given by
$$
\frac{m_u}{m_d}\simeq 3 \frac{m_s}{m_c} \simeq \frac{3}{2}\beta_d
\ , \eqno(4.3)
$$
so that the ratios $m_u/m_d$ and $m_s/m_c$ can be fitted
independently of $m_t/m_c$
(i.e., $K/k$) by adjusting the parameter $\beta_d$.

When we take $b_d=-1.0$ and $\beta_d=-18^\circ$
(and $k=10$ and $K/k=50$),
we can obtain
reasonable quark mass values:
$$
\begin{array}{lll}
m_u(\Lambda_W)=0.00234 \ {\rm GeV} \ , & m_c(\Lambda_W)=0.610 \ {\rm GeV}\ ,
& m_t(\Lambda_W)=166 \ {\rm GeV}\ , \\
m_d(\Lambda_W)=0.00475 \ {\rm GeV}\ ,  & m_s(\Lambda_W)=0.0923 \ {\rm GeV} \ ,
& m_b(\Lambda_W)=3.01 \ {\rm GeV}\ , \\
\end{array} \eqno(4.4)
$$
where we have taken $m_0(\Lambda_W)=288$ GeV to have
$m_t(\Lambda_W)=166$ GeV.

So far, except for (4.4), we have discussed only quark mass ratios and not
the absolute values, because the ratios are comparatively insensitive
to the evolution from $\mu=m_0 K$ to $\mu=m_0$.
The common value $m_0(\Lambda_W)=288$ GeV does not give the absolute
magnitudes of the charged lepton masses,
$(k/K)m_0=m_\tau+m_\mu+m_e$.
We find
$$
\left. \frac{(m_0 k/K)_q}{(m_0 k/K)_\ell} \right|_{\mu=\Lambda_W}
=3.05 \ , \eqno(4.5)
$$
where $(m_0 k/K)_{q(\ell)}$ denotes the value of $m_0 k/K$
in the quark (lepton) sector.
It is not likely that the factor 3.1  comes
only from the evolution from $\mu=m_0 K$ to the
present scale $\mu=\Lambda_W$.
Since we consider the case where the parameters $m_0$ and $k$
(i.e., $m_L$ and $m_R$) are universal for all flavors
$f=u,d,\nu,e$, the discrepancy (4.5)
should come from the difference in $K$ between the quark- and
lepton-sectors, i.e., $K_q\neq K_\ell$.
Although it is possible that the coupling constants of
the colored heavy fermions with Higgs bosons which
generate the democratic-type matrix (2.2) are smaller than
that of the colorless heavy fermions by a factor 1/3,
i.e., $K_\ell/K_q=3$,  we do not
discuss the origin of $K_\ell/K_q=3$ in the present paper.
In the present model, we practically consider that $m_L$ and $m_R$ are
universal for quarks and leptons, while $M_F$ are not so, and
$K_u=K_d\equiv K_q\neq K_\nu=K_e\equiv K_\ell$.
Hereafter, we denote $K_q$ simply as $K$.

Similarly, with the same parameter values as in (4.4),
the heavy quark masses  are given as follows:
$$
\begin{array}{lll}
m_4^u(\Lambda_W)=1.66 \ {\rm TeV} \ , & m_5^u(\Lambda_W)=144 \ {\rm TeV}\ ,
& m_6^u(\Lambda_W)=144 \ {\rm TeV}\ , \\
m_4^d(\Lambda_W)=144 \ {\rm TeV}\ , & m_5^d(\Lambda_W)=144 \ {\rm TeV}\ ,
& m_6^d(\Lambda_W)=298 \ {\rm TeV}\ . \\
\end{array} \eqno(4.6)
$$
These numerical results are also obtained from the approximate relations
for $b_u=-1/3$ and $b_d=-1$:
$$
m_4^u \simeq (k/\sqrt{3}) m_0 \ , \ \ \
m_5^u \simeq m_6^u \simeq K m_0\ ,
\eqno(4.7)
$$
$$
m_4^d\simeq m_5^d\simeq K m_0\ , \ \ \
m_6^d \simeq 2\sqrt{1+3\beta_d^2/4} K m_0\ .
\eqno(4.8)
$$
Note that the fourth up-quark $u_4$ becomes considerably
lighter than the other heavy quarks, at the cost of
the enhancingthe top-quark mass.
The absolute magnitudes the heavy quark masses in (4.6)
should not be taken solidly,
because they depend on both  $k$ and $K$.
We have chosen $K/k=50$ in order to fit $m_t/m_c$, but the choice
$k=10$ was only a trial choice, because the predictions for light
fermions (quarks and leptons) are insensitive to the value of $k$.
Only constraint on the value $k$ comes from the relation
$k \sim m(W_R^\pm)/m(W_L^\pm)$.
The present lower bound of the right-handed weak boson mass $m(W_R)$
is given in Ref.~[15], so that we cannot choose too small value of $k$.
Since $m_4^u$ is of the order of $k m_0$, as seen in (4.7),
we can expect to observe the fourth up-quark
at the energy scale where the right-handed weak bosons
$W_R$ are observed.


\vglue.2in

\noindent
{\large\bf 5.\  KM matrix parameters}

In the present model,
the parameter fitting for five quark-mass ratios and four KM matrix
parameters is done by five parameters, $k/K$ (not $k$ and $K$),
$b_u$, $\beta_u$, $b_d$ and $\beta_d$.
When we adopt the ansatz of ``maximal top-quark-mass enhancement",
we have fixed the parameters $b_u$ and $\beta_u$ to $b_u=-1/3$
and $\beta_u=0$, and the remaining adjustable parameters are
$k/K$, $b_d$ and $\beta_d$.
We have pointed out that the relation between up-quark mass ratio $m_u/m_c$
and $m_e/m_\mu$, (1.4), is satisfied independently of these parameters
for the case $b_u\simeq -1/3$.
The parameter $K/k$ was fixed to $K/k=50$ from the observed up-quark
mass ratio $m_t/m_c$, see Fig.~2.
In the previous section, we have shown that the remaining two
parameter $b_d$ and $\beta_d$ can be fitted to three observed quark mass
ratios $m_d/m_s$, $m_s/m_b$ and $m_u/m_d$ reasonably (see Fig.~4).
Then, our final task in the present phenomenological study is
to check whether these parameter values can also give reasonable
predictions for the four KM matrix parameters.

The KM matrix $V$ is given by $V = U_u U_d^\dagger$,
where $U_q$ ($q=u,d$) are the unitary matrices to diagonalize
the light fermion mass matrices $M_f M_f^\dagger$,
where $M_f\equiv M'_{11}$ ($f=u,d$) defined by (3.2).
Unfortunately, our parameter values $K/k\simeq 50$, $b_d\simeq -1$
and $\beta_d\simeq -18^\circ$  give rise to the KM matrix parameters
far away  from the observed values [13].
Therefore, we must slightly modify our model.

So far, we have assumed that the matrices $m_L$ and $m_R$
are universal for up- and down-sectors.
However, in the present section, let us distinguish the
matrix $m_{L}$ in the up-quark sector,
$m^u_{L}=m_0 Z_u$,
from that in down-quark sector, $m_L^d=m_0 Z_d$.
We assume that $Z_u$ and $Z_d$ are given by
$Z_q=P_q Z $ ($q=u,d$), where $Z$ is given
by (2.3) and (2.4), and $P_q$ are phase matrices.
(It is not essential whether we also assume a similar
modification on $m_R$ or not, because the KM matrix is
related only to the family mixing among the left-handed fields.)
Such a modification does not change our predictions
on the fermion masses in Sects.~3 and 4, while
the KM matrix $V$ is changed into the following
expression:
$$
V = U_u P U_d^\dagger \ , \eqno(5.1)
$$
where $U_q$ ($q=u,d$) are unitary matrices to diagonalize
the unchanged matrices $M_f M_f^\dagger$
(i.e., in the case of $P_u=P_d={\bf 1}$), and
$P=P_u P_d^\dagger$.
In general, the phase matrix $P$ can have two independent
phase parameters such as
$P={\rm diag}(1,e^{i\delta_2},e^{i\delta_3})$.
However, since we do not want more adjustable parameters,
we examine a simpler ansatz that the phase matrix $P$ is real, i.e.,
$\delta_i=0$ or $\pi$.
Thus, we keep three adjustable parameters,
$k/K$, $b_d$ and $\beta_d$, at the cost of putting
the additional ansatz on $P$.

As a result, we find that only for the case
$$
P={\rm diag}(1, 1, -1) \ , \eqno(5.2)
$$
we can obtain  reasonable
values of $|V_{us}|$, $|V_{cb}|$ and $|V_{ub}|$.
We show $|V_{us}|$, $|V_{cb}|$ and $|V_{ub}|$ versus $\beta_d$
in Fig.~5.
The same parameter values as in (4.4), $K/k=50$, $b_d=-1$ and
$\beta_d=-18^\circ$, give reasonable predictions
$$
|V_{us}|=0.220 \ , \ \  |V_{cb}|=0.0598 \ , \ \  |V_{ub}|=0.00330 , \ \
|V_{td}|=0.0155 ,
$$
$$
J=-3.18 \times 10^{-5} \ ,
\eqno(5.3)
$$
where $J$ is the rephasing invariant [16]
$J={\rm Im} (V_{cb}V_{us}V^*_{cs}V^*_{ub})$.
Although the origin of the phase inversion $P={\rm diag}(1,1,-1)$
is not clear and the predicted value of $V_{cb}$ is somewhat large,
it is a noticeable feature of the present model
that the parameters which were fixed by the observed
quark-mass ratios can roughly give reasonable predictions for all
the KM matrix parameters.

\vglue.2in

\noindent
{\large\bf 6.\  Conclusions}

In conclusion, we have demonstrated that the seesaw-type mass matrix (2.1)
with $M_F$ given by (2.2) can give top-quark-mass enhancement without
assuming any parameters with hierarchically different values between
$M_U$ and $M_D$, i.e., with  $b_u\simeq -1/3$
and $b_d\simeq -1$.
The enhancement $m_t/m_b\gg 1$ comes from the fact that the democratic
part $X$ in the inverse matrix $M_F^{-1}$  in (1.2), is
enhanced as to $b_f\rightarrow -1/3$
because $|a_f|\rightarrow \infty$ in the limit as seen in (1.8).
On the other hand, the result $m_u\sim m_d$ comes from the feature
that the democratic-type mass matrix can give rise to a large mass
only to the third family, i.e., the effect of $|a_u|\rightarrow\infty$
contributes mainly to $m_t$.

In the present model,
the parameter fitting for the five quark mass ratios and
the four KM matrix parameters has been done
by five parameters $k/K$ (not $k$ and $K$ separately),
$b_u$, $\beta_u$, $b_d$ and $\beta_d$.
(The parameters $z_i$ were fixed by charged lepton masses.)
When we adopt the ansatz of ``maximal top-quark-mass enhancement",
the parameters $b_u$ and $\beta_u$ are fixed to $b_u=-1/3$
and $\beta_u=0$, and the remaining adjustable parameters are
$k/K$, $b_d$ and $\beta_d$.
The parameter $K/k$ is then fixed by the observed up-quark-mass
ratio $m_t/m_c$ to be $K/k=50$ .
The remaining two parameters $b_d$ and $\beta_d$ are
then free parameters by which
four quark mass ratios $m_u/m_c$, $m_d/m_s$,
$m_s/m_b$ and $m_u/m_d$, and four KM parameters are fitted.
As shown in Sects.~4 and 5, by choosing $b_d\simeq -1$ and
$\beta_d \simeq -18^\circ$, we have obtained reasonable fitting for the
quark-mass ratios, and also for the KM matrix parameters
with the ansatz (5.2).

A few remarks are in order.

In the present model, flavor changing neutral currents
(FCNC) can, in principle,  appear.
However, the FCNC due to the SU(2)$_L$ (SU(2)$_R$) doublet Higgs boson
exchange through $f$-$F$ mixing are highly suppressed
by a GIM-like mechanism [17].
The FCNC due to the $Z$-boson exchange through $f$-$F$ mixing are also
suppressed because the effective coupling constants are order of
$1/K$ (we can find that those are of the order of $10^{-8}$ in the case
of $k=10$), so that the FCNC rare decay modes are suppressed by
$10^{-16}$.

The $CP$ violating phases come only from
the heavy fermion mass matrix $M_F$, i.e., from the parameter $\beta_f$.
In the up-quark sector, the parameter $\beta_u$ must be $\beta_u=0$,
because the top quark mass enhancement becomes mild
when $\beta_u\neq 0$.
On the other hand, if $\beta_d=0$, we cannot fit down-quark mass ratios
$m_d/m_s$ and $m_s/m_b$ for any values of $k/K$ and $b_d$.
We must choose a sizable value of $\beta_d$.
Thus, in our model, the $CP$ violating phase in quarks comes only
from the down-quark sector $M_D$.

In the present paper, we have discussed a seesaw mass matrix model
with the form of $M_F=m_0 K O_F$ given by (2.2).
As far as the phenomenological predictions are concerned,
we can choose other family-basis, for example,
a rather simple form of $O_F$
$$
O_F={\bf 1} +3 b_f e^{i\beta_f} {\rm diag}(0, 0, 1) \ , \eqno(6.1)
$$
instead of the democratic form (2.2).
However, in order to obtain reasonable predictions of quark mass
ratios and KM matrix parameters,
the matrix $Z$ cannot be a diagonal form such as in (2.3),
and it must be given by
$$
Z=\frac{1}{6} \left(
\begin{array}{ccc}
3(z_2+z_1) &  -\sqrt{3}(z_2-z_1) & -\sqrt{6}(z_2-z_1) \\
 -\sqrt{3}(z_2-z_1) & 4z_3+z_2+z_1 & -\sqrt{2}(2z_3 -z_2-z_1) \\
-\sqrt{6}(z_2-z_1) &  -\sqrt{2}(2z_3 -z_2-z_1) & 2(z_3+z_2+z_1) \\
\end{array} \right) \ ,\eqno(6.2)
$$
where $z_i$ are given by (2.4).
Which family basis is reasonable is not essential as far as
we discuss only the
fermion masses and KM mixing parameters, but it will
become important for model-building.

We believe that our phenomenological mass-matrix model is worth
serious attention, not only because it has fewer adjustable
parameters than conventional models do, but also because it gives
$m_t \gg m_b$ and $m_u \sim m_d$
simultaneously despite its  ``almost"
up-down symmetric mass matrices (i.e., $b_u/b_d$ is not so
large as $m_t/m_b$).

\newpage

\centerline{\bf Acknowledgments}

The authors would like to express their sincere thanks Professors
R.~Mohapatra for a careful reading of the original manuscript
and providing valuable comments
and Professor M.~Tanimoto for helpful discussions.
The authors are also indebted to Professors K.~Matumoto
for his encouragement and his stimulating comments.
This work was supported by the Grant-in-Aid for Scientific Research,
Ministry of Education, Science and Culture, Japan (No.06640407).


\vglue.4in

\centerline{\bf Appendix: Diagonalization of $2n\times 2n$ matrix}

The transformation of $2n\times 2n$ matrix
$$
M = \left(
\begin{array}{cc}
M_{11} & M_{12} \\
M_{21} & M_{22}
\end{array} \right) \ \eqno(A.1)
$$
into
$$
M^{'} \equiv \left(
\begin{array}{cc}
M_{11}^{'} & 0 \\
0 & M_{22}^{'}
\end{array} \right)  \
\eqno(A.2)
$$
is done by two $2n\times 2n$ unitary matrices,
$$ U_L = \left(
\begin{array}{cc}
(1+\rho_L\rho_L^\dagger)^{-1/2} & (1+\rho_L\rho_L^\dagger)^{-1/2}\rho_L \\
-(1+\rho_L^\dagger\rho_L)^{-1/2}\rho_L^\dagger &
(1+\rho_L^\dagger\rho_L)^{-1/2}
\end{array} \right) \  \eqno(A.3)
$$
and $U_R$ with $L\leftrightarrow R$ in (A.3) as
$$
M'= U_L M U_R^\dagger \ , \eqno(A.4)
$$
where $M_{ij}$, $M'_{ij}$, $\rho_L$, $\rho_R$ are $n\times n$ matrices.

The conditions $M'_{12}=0$ and $M'_{21}=0$ lead to the relations
$$
M_{12}-M_{11}\rho_R+\rho_LM_{22}
-\rho_LM_{21}\rho_R=0 \ , \eqno(A.5)
$$
and
$$
M_{21}+M_{22}\rho_R^\dagger-\rho_L^\dagger M_{11}
-\rho_L^\dagger M_{12}\rho_R^\dagger=0 \ , \eqno(A.6)
$$
respectively, which lead to
$$
\rho_R =
(M_{11}+\rho_LM_{21})^{-1}
(M_{12}+\rho_LM_{22}) \ , \eqno(A.7)
$$
$$
\rho_R =
(M_{21}^\dagger-M_{11}^\dagger\rho_L)
(M_{12}^\dagger\rho_L-M_{22}^\dagger)^{-1}
\ . \eqno(A.8)
$$
By eliminating $\rho_R$ from (A.7) and (A.8),
we obtain
$$
(M_{12}+\rho_LM_{22})
(M_{12}^\dagger\rho_L-M_{22}^\dagger)
=
(M_{11}+\rho_LM_{21})
(M_{21}^\dagger-M_{11}^\dagger\rho_L)
\ , \eqno(A.9)
$$
or
$$
M_{11}M_{21}^\dagger+M_{12}M_{22}^\dagger
-(M_{11}M_{11}^\dagger
+M_{12}M_{12}^\dagger)
\rho_L
$$
$$
+\rho_L(M_{21}M_{21}^\dagger
+M_{22}M_{22}^\dagger)
-\rho_L(
M_{21}M_{11}^\dagger+M_{22}M_{12}^\dagger)
\rho_L= 0 \ , \eqno(A.10)
$$
Similarly, we obtain the relation
$$
M_{11}^\dagger M_{12}+M_{21}^\dagger M_{22}
-(M_{11}^\dagger M_{11}+M_{21}^\dagger M_{21})
\rho_R
$$
$$
+\rho_R(M_{12}^\dagger M_{12}+
M_{22}^\dagger M_{22})-
\rho_R(
M_{12}^\dagger M_{11}+
M_{22}^\dagger M_{21})\rho_R
= 0 \ . \eqno(A.11)
$$

Eliminating $M_{22}$ from (A.5) and (A.6),
we obtain
$$
\rho_LM_{22}\rho_R^\dagger =
(M_{11}\rho_R+\rho_LM_{21}\rho_R-M_{12})\rho_R^\dagger
$$
$$
= \rho_L(\rho_L^\dagger M_{11}+
\rho_L^\dagger M_{12}\rho_R^\dagger-M_{21})
\ , \eqno(A.12)
$$
so that
$$
(1+\rho_L\rho_L^\dagger)(M_{11}+M_{12}\rho_R^\dagger)
=(M_{11}+\rho_LM_{21})(1+\rho_R\rho_R^\dagger) \ . \eqno(A.13)
$$
Similarly, eliminating $M_{11}$ from (A.5) and (A.6),
we obtain
$$
\rho_L^\dagger M_{11}\rho_R =
\rho_L^\dagger(M_{12}+\rho_LM_{22}-\rho_LM_{21}\rho_R)
$$
$$
= (M_{21}+M_{22}\rho_R^\dagger-
\rho_L^\dagger M_{12}\rho_R^\dagger)\rho_R
\ , \eqno(A.14)
$$
so that
$$
(1+\rho_L^\dagger\rho_L)(M_{22}-M_{21}\rho_R)=
(M_{22}-\rho_L^\dagger M_{12})(1+\rho_R^\dagger\rho_R)
\ . \eqno(A.15)
$$
By using the relations (A.13) and (A.15),
we obtain
$$
M_{11}^{'} =
(1+\rho_L\rho_L^\dagger)^{-1/2}(M_{11}+
M_{12}\rho_R^\dagger+\rho_LM_{21}+\rho_L
M_{22}\rho_R^\dagger)
(1+\rho_R\rho_R^\dagger)^{-1/2}
$$
$$
=
(1+\rho_L\rho_L^\dagger)^{-1/2}
(M_{11}+\rho_LM_{21})
(1+\rho_R\rho_R^\dagger)^{+1/2} \  \eqno(A.16)
$$
$$
=
(1+\rho_L\rho_L^\dagger)^{+1/2}
(M_{11}+M_{12}\rho_R^\dagger)(1+\rho_R\rho_R^\dagger)^{-1/2}
\ , \eqno(A.17)
$$
$$
M_{22}^{'} =
(1+\rho_L^\dagger\rho_L)^{-1/2}
(\rho_L^\dagger M_{11}\rho_R-
\rho_L^\dagger M_{12}-
M_{21}\rho_R
+M_{22})
(1+\rho_R^\dagger\rho_R)^{-1/2}
$$
$$
=
(1+\rho_L^\dagger\rho_L)^{+1/2}
(M_{22}-M_{21}\rho_R)(1+\rho_R^\dagger\rho_R)^{-1/2}
\ \eqno(A.18)
$$
$$
=
(1+\rho_L^\dagger\rho_L)^{-1/2}
(M_{22}-\rho_L^\dagger M_{12})
(1+\rho_R^\dagger\rho_R)^{+1/2}
\ . \eqno(A.19)
$$
The matrices $\rho_L$ and $\rho_R$ are obtained as
solutions of the equations (A.10) and (A.11),
respectively.
When the $2n\times 2n$ mass matrix $M$ (A.1) is Hermitian,
we can set $\rho_L =\rho_R \equiv \rho$, so that
the calculation becomes easier.

When $M$ is not Hermitian,
instead of the $n\times n$ mass matrices $M'_{11}$
and $M'_{22}$, the diagonalization is
done for the following Hermitian matrices
$$
H_1 \equiv M'_{11} M_{11}^{\prime \dagger} =
(1+\rho_L\rho_L^\dagger)^{-1/2}\widetilde{H}_1
(1+\rho_L\rho_L^\dagger)^{+1/2} \ , \eqno(A.20)
$$
$$
H_2 \equiv M'_{22} M_{22}^{\prime \dagger} =
(1+\rho_L^\dagger\rho_L)^{+1/2}\widetilde{H}_2
(1+\rho_L^\dagger\rho_L)^{-1/2} \ , \eqno(A.21)
$$
where
$$
\widetilde{H}_1 =(M_{11} +\rho_L M_{21})
(M_{11}^\dagger +\rho_R M_{12}^\dagger) \ ,
\eqno(A.22)
$$
$$
\widetilde{H}_2 =(M_{22} - M_{21} \rho_R)
(M_{22}^\dagger -M_{12}^\dagger \rho_L) \ .
\eqno(A.23)
$$

We are interested in the diagonalization of (A.22).
By using (A.5), we can rewrite (A.22) into
$$
\widetilde{H}_1= A + \rho_L B \ , \eqno(A.24)
$$
where
$$
A=M_{11} M_{11}^\dagger + M_{12} M_{12}^\dagger \ ,
\eqno(A.25)
$$
$$
B=M_{21} M_{11}^\dagger + M_{22} M_{12}^\dagger \ .
\eqno(A.26)
$$
By eliminating $\rho_L$ from (A.10) and (A.24),
we find that the matrix $\widetilde{H}_1$ satisfies
the following equations
$$
\widetilde{H}_1^2 - \widetilde{H}_1 (A+B^{-1}DB)
+A B^{-1} DB - CB =0 \ , \eqno(A.27)
$$
where
$$
C=M_{11} M_{21}^\dagger + M_{12} M_{22}^\dagger \ ,
\eqno(A.28)
$$
$$
D=M_{21} M_{21}^\dagger + M_{22} M_{22}^\dagger \ .
\eqno(A.29)
$$

\newpage

\vglue.3in
\newcounter{0000}
\centerline{\bf References and Footnotes}
\begin{list}
{[~\arabic{0000}~]}{\usecounter{0000}
\labelwidth=0.8cm\labelsep=.1cm\setlength{\leftmargin=0.7cm}
{\rightmargin=.2cm}}

\item The seesaw mechanism has originally proposed for the purpose of
explaining why neutrino masses are so invisibly small:
T.~Yanagida, in {\it Proc. Workshop of the Unified Theory and
Baryon Number in the Universe}, edited by A.~Sawada and A.~Sugamoto
(KEK, 1979);
M.~Gell-Mann, P.~Rammond and R.~Slansky, in {\it Supergravity},
edited by P.~van Nieuwenhuizen and D.~Z.~Freedman (North-Holland,
1979);
R.~Mohapatra and G.~Senjanovic, Phys.~Rev.~Lett.
{\bf 44} (1980)  912.

\item For  applications of the seesaw mechanism
to the quark mass matrix, see, for example,
Z.~G.~Berezhiani, Phys.~Lett. {\bf 129B} (1983)  99;
Phys.~Lett. {\bf 150B} (1985)  177;
D.~Chang and R.~N.~Mohapatra, Phys.~Rev.~Lett. {\bf 58}
(1987) 1600;
A.~Davidson and K.~C.~Wali, Phys.~Rev.~Lett. {\bf 59} (1987)  393;
S.~Rajpoot, Mod.~Phys.~Lett. {\bf A2} (1987)  307;
Phys.~Lett. {\bf 191B} (1987)  122; Phys.~Rev. {\bf D36} (1987)  1479;
K.~B.~Babu and R.~N.~Mohapatra, Phys.~Rev.~Lett. {\bf 62}  (1989) 1079;
Phys.~Rev. {\bf D41} (1990)  1286;
S.~Ranfone, Phys.~Rev. {\bf D42} (1990)  3819;
A.~Davidson, S.~Ranfone and K.~C.~Wali, Phys.~Rev.
{\bf D41} (1990)  208;
I.~Sogami and T.~Shinohara, Prog.~Theor.~Phys. {\bf 66}  (1991)  1031;
Phys.~Rev. {\bf D47}  (1993)  2905;
Z.~G.~Berezhiani and R.~Rattazzi, Phys.~Lett. {\bf B279}  (1992)  124;
P.~Cho, Phys.~Rev. {\bf D48} (1994)  5331;
A.~Davidson, L.~Michel, M.~L,~Sage and  K.~C.~Wali, Phys.~Rev.
{\bf D49} (1994)  1378;
W.~A.~Ponce, A.~Zepeda and R.~G.~Lozano, Phys.~Rev. {\bf D49}
(1994) 4954.
\item Y.~Koide, Phys.~Rev. {\bf D49} (1994)  2638.
\item H.~Harari, H.~Haut and J.~Weyers, Phys.~Lett. {\bf B78}
(1978) 459;
T.~Goldman, in {\it Gauge Theories, Massive Neutrinos and
Proton Decays}, edited by A.~Perlumutter (Plenum Press, New York,
1981), p.111;
T.~Goldman and G.~J.~Stephenson,~Jr., Phys.~Rev. {\bf D24}
(1981) 236;
Y.~Koide, Phys.~Rev.~Lett. {\bf 47} (1981) 1241; Phys.~Rev.
{\bf D28} (1983) 252; {\bf 39} (1989) 1391;
C.~Jarlskog, in {\it Proceedings of the International Symposium on
Production and Decays of Heavy Hadrons}, Heidelberg, Germany, 1986
edited by K.~R.~Schubert and R. Waldi (DESY, Hamburg), 1986, p.331;
P.~Kaus, S.~Meshkov, Mod.~Phys.~Lett. {\bf A3} (1988) 1251;
Phys.~Rev. {\bf D42} (1990) 1863;
L.~Lavoura, Phys.~Lett. {\bf B228} (1989) 245;
M.~Tanimoto, Phys.~Rev. {\bf D41} (1990) 1586;
H.~Fritzsch and J.~Plankl, Phys.~Lett. {\bf B237} (1990) 451;
Y.~Nambu, in {\it Proceedings of the International Workshop on
Electroweak Symmetry Breaking}, Hiroshima, Japan, (World
Scientific, Singapore, 1992), p.1.
\item Y.~Koide, Mod.~Phys.~Lett. {\bf A8} (1993)  2071.
\item H.~Fusaoka and Y.~Koide, Mod. Phys.~Lett. {\bf A10} (1995)  289.
\item M.~Kobayashi and T.~Maskawa, Prog.~Theor.~Phys. {\bf 49} (1973)  652.
\item CDF collaboration, F.~Abe {\it{et al.}}, Phys.~Rev.~Lett.
{\bf 73} (1994)  225; Phys.~Rev. {\bf D50} (1994)  2966.
\item A seesaw neutrino mass matrix with a democratic-type matrix
$M_F$ was first discussed by Goldman and Stephenson in ref.~[4].
Their interest, however, was not in the quark mass matrix,
and their matrix $M_F$ does not include the unit
matrix part, which is different from (2.2).
\item H.~Harari, H.~Haut and J.~Weyers, Phys.~Lett. {\bf B78} (1978) 459.
\item P.~Kaus, S.~Meshkov, Mod.~Phys.~Lett. {\bf A3} (1988) 1251;
Phys.~Rev. {\bf D42} (1990) 1863;
Y.~Nambu, in {\it Proceedings of the International Workshop on
Electroweak Symmetry Breaking}, Hiroshima, Japan, (World
Scientific, Singapore, 1992), p.1.
\item Y.~Koide, Phys.~Rev.~Lett. {\bf 47} (1981) 1241.
\item Particle data group, Phys.~Rev. {\bf D50} (1994) 1173.
\item For the light quark masses $m_q(\mu)$,
we have used the values at $\mu=1$ GeV,
$m_u=5.6\pm 1.1$ MeV, $m_d=9.9\pm 1.1$ MeV and $m_s=199\pm 33$ MeV:
C.~A.Dominquez and E.~de Rafael, Annals of Physics {\bf 174} (1987)  372.
However, the absolute values of light quark masses should not be taken
solidly because they depend on models.
For $m_c(\mu)$ and $m_b(\mu)$,
we have used the value $m_c(m_c)=1.26\pm 0.02$ GeV by
S.~Narison, Phys.~Lett. {\bf B216} (1989) 191,
and the value $m_b(m_b)=4.72\pm 0.05$ GeV by
C.~A.~Dominquez and N.~Paver, Phys.~Lett. {\bf B293} (1992)  197.
For the top quark mass, we have used
$M_t^{pole}=174\pm 10 ^{+13}_{-12}$ GeV
from the CDF experiment [8].
\item P.~Langacker and S.~U.~Sankar, Phys.~Rev. {\bf D40} (1989) 1569.
\item C.~Jarlskog, Phys.~Rev.~Lett. {\bf 55} (1985)  1839;
O.~W.~Greenberg, Phys.~Rev. {\bf D32} (1985)  1841;
I.~Dunietz, O.~W.~Greenberg, and D.-d.~Wu, Phys.~Rev.~Lett. {\bf 55}
 (1985) 2935;
C.~Hamzaoui and A.~Barroso, Phys.~Lett. {\bf 154B} (1985) 202;
D.-d.~Wu, Phys.~Rev. {\bf D33} (1986)  860.
\item S.~L.~Glashow, J.~Iliopulos and L.~Maiani, Phys.~Rev.
{\bf D2} (1970) 1285.
\end{list}

\newpage

\begin{center}
{\bf Figure Captions}
\end{center}

Fig.~1. Masses  $m_i^f$ ($i=1,2,3$) versus $b_f$ for the case of
$k=10$ and  $K/k=50$.
The solid and broken lines denote for the cases of $\beta_f=0$
and $\beta_f=-20^\circ$, respectively.
The parameters $k$ and $K$ are defined by (2.1).
The figure should be taken as that for the quark mass ratios.
For the absolute value of quark masses, see a comment on (4.5)
in the text.
\vglue.1in

Fig.~2.  Mass ratio $m_t/m_c$ versus $K/k$ for $k=10$.
The curves (a) -- (d)
denote the cases (a) $\Delta b_u=0$,
(b) $\Delta b_u=+1.00\times 10^{-3}$ and $\Delta b_u=-0.980\times 10^{-3}$,
(c) $\Delta b_u=+3.88\times 10^{-3}$ and $\Delta b_u=-3.62\times 10^{-3}$,
(d) $\Delta b_u=+10.0\times 10^{-3}$ and $\Delta b_u=-8.53\times 10^{-3}$.
The horizontal lines denote the experimental values $(m_t/m_c)_{exp}
=271\pm 46$.

\vglue.1in

Fig.~3. Top quark mass $m_t$ in unit of $m_0$ versus $K/k$ for
$k=10$.
The curves (a) -- (d)
denote the cases (a) $\Delta b_u=0$,
(b) $\Delta b_u=+1.00\times 10^{-3}$ and $\Delta b_u=-0.980\times 10^{-3}$,
(c$_+$) $\Delta b_u=+3.88\times 10^{-3}$,
(c$_-$) $\Delta b_u=-3.62\times 10^{-3}$,
(d$_+$) $\Delta b_u=+10.0\times 10^{-3}$,
and (d$_-$) $\Delta b_u=-8.53\times 10^{-3}$.

\vglue.1in

Fig.~4. Mass ratios $m_s/m_d$ and $m_b/m_s$ versus $\beta_d$
for $b_d=-0.90$ (a dotted line), $b_d=-1.0$ (a solid line)
and $b_d=-1.1$ (a broken line) in the case of
 $k=10$ and $K/k=50$.

\vglue.1in

Fig.~5. Kobayashi-Maskawa matrix elements $|V_{us}|$, $|V_{cb}|$
and $|V_{ub}|$ versus $\beta_d$ in the case of
 $k=10$, $K/k=50$, $b_u=-1/3$ and $\beta_u=0$ .


\end{document}